\documentclass[pop,numberedappendix,iop]{aeb_emulateapj_2015}
\usepackage{amsmath,cases}
\usepackage{color}
\usepackage[mathscr]{euscript}
\usepackage{ulem}
\usepackage{hyperref}

\begin{document}

\author
{
Mohamad Shalaby\altaffilmark{1,2}, 
Avery E. Broderick\altaffilmark{1,3}, 
Philip Chang\altaffilmark{4},
\\
Christoph Pfrommer\altaffilmark{5}, 
Astrid Lamberts\altaffilmark{6}
and
Ewald Puchwein\altaffilmark{7}
}

\altaffiltext{1}{Perimeter Institute for Theoretical Physics, 31 Caroline Street North, Waterloo, ON, N2L 2Y5, Canada}
\altaffiltext{2}{Department of Astronomy and Astrophysics, University of Chicago, 5640 S Ellis Ave, Chicago, IL 60637, United States}
\altaffiltext{3}{Department of Physics and Astronomy, University of Waterloo, 200 University Avenue West, Waterloo, ON, N2L 3G1, Canada}
\altaffiltext{4}{Department of Physics, University of Wisconsin-Milwaukee, 1900 E. Kenwood Boulevard, Milwaukee, WI 53211, USA}
\altaffiltext{5}{Leibniz-Institut f{\"u}r Astrophysik Potsdam (AIP), An der Sternwarte 16, 14482 Potsdam, Germany}
\altaffiltext{6}{Theoretical Astrophysics, California Institute of Technology, Pasadena, CA 91125, USA}
\altaffiltext{7}{Institute of Astronomy and Kavli Institute for Cosmology, University of Cambridge, Madingley Road, Cambridge, CB3 0HA, UK}

\email{mshalaby@live.ca}

\title{Growth of beam-plasma instabilities in the presence of background inhomogeneity}

\begin{abstract}

We explore how  inhomogeneity in the background plasma number density alters the growth of electrostatic unstable wavemodes of beam plasma systems.
This is particularly interesting for blazar-driven beam-plasma instabilities, which may be suppressed by inhomogeneities in the intergalactic medium as was recently claimed in the literature.
Using high resolution Particle-In-Cell simulations with the SHARP code, we show that the growth of the instability is local, i.e., regions with almost homogeneous background density will support the growth of the Langmuir waves, even when they are separated by strongly inhomogeneous regions, resulting in an overall slower growth of the instability.
We also show that if the background density is continuously varying, the growth rate of the instability is lower; though in all cases, the system remains within the linear regime longer and the instability is not extinguished.
In all cases, the beam loses approximately the same fraction of its initial kinetic energy in comparison to the uniform case at non-linear saturation.
Thus, inhomogeneities in the intergalactic medium are unlikely to suppress the growth of blazar-driven beam-plasma instabilities.

\end{abstract}

\section{Introduction}

The majority of astrophysical plasmas are cold ($k_B T \ll m_e c^2$)
and collisionless.
Plasmas that contain non-thermal (relativistic or non-relativistic) sub-populations are subject to strong beam-plasma instabilities that can redistribute the energy in nonthermal populations.
Of particular interest are the beam-plasma instabilities due to the propagation of the electron-positron pairs driven by the TeV emission of blazars in the intergalactic medium (IGM)~\citep{blazarI,blazarII,blazarIII,Schlickeiser+12,Schlickeiser+13,Vafin+2018}.
These could lead to, e.g., a preferential heating of low density regions of the IGM~\citep{Puchwein+2011,patchy-heating+2015}.

Linear and quasi-linear analyses of the beam-plasma instabilities assume that the background plasma is spatially uniform both when the growth rates for oblique wavemodes are computed by, e.g.,~\citet{Bret-2010-PRE,Bret-2010-POP,linear-paper} and also when higher order perturbative calculations are used to assess the nonlinear effects on the linear growth rates~\citep{nonlinear-paper}.
However, as pointed out by \citet{Breizman+Inhomo01,Miniati-Elyiv-2013}, the background inhomogeneity may be particularly problematic.
In particular, it may completely suppress the effect of the instabilities for blazar-driven beam-plasma instabilities in the IGM.

A relevant characterization of the inhomogeneity in the beam-plasma system is the inhomogeneity scale length along the beam direction, $\lambda^{\rm inh}_{\parallel}$.
This is a measure of the spatial scale over which the number density, $n$,  changes significantly along the beam direction and is typically defined as $\lambda^{\rm inh}_{\parallel} \equiv  |n/  (\vec{\nabla} n )_{\parallel}  |$.
The inhomogeneity scale for the IGM at different redshifts, $z$, was computed in \citet{Miniati-Elyiv-2013} using cosmological simulations.
They find that at mean density
$\lambda^{\rm inh}_{\parallel} \sim 25$, $100$, and $400$ kpc for $z=3$, $1$ and $0$, respectively.
The distance travelled by the blazar-induced pair-beams is estimated to be about $1$ kpc in one growth time, i.e., one  $e$-folding  \citep[][]{Miniati-Elyiv-2013}.
Therefore, the pair beams experience a slowly varying IGM number density.

In the presence of background plasma inhomogeneity, there are two timescales that are important for determining the effect of the inhomogeneity on the growth of the unstable wavemodes.
First, the timescale for the growing wave to respond to the inhomogeneity, $\tau_{\rm inh}$, and, second, the timescale over which the growth occurs in the linear regime, $\tau_{\rm g}$, i.e., the timescale after which nonlinear effects become important.
The growth rates for the homogeneous background plasma are applicable if 
\begin{eqnarray}
\label{eq:growth-condition}
\tau_{\rm inh} \gg \tau_{\rm g}.
\end{eqnarray}
When this condition is violated, i.e., $\tau_{\rm inh} \lesssim \tau_{\rm g}$, the growth of wavemodes can still occur but at lower rates \citep{Breizman+Inhomo01}.
However, the degree of this suppression is uncertain.
\citet{Miniati-Elyiv-2013} assumed that the unstable wavemodes are completely suppressed when the condition in Equation~\eqref{eq:growth-condition} is violated.
Contrarily, \citet{Breizman+Inhomo01} claim that the beam loses only a negligible amount of its initial kinetic energy when this condition is violated.

In this work, we use high resolution one-dimensional particle-in-cell (PIC) simulations to study the effect of number density inhomogeneity.
We use the SHARP code~\citep{sharp}, which provides an excellent control over typical numerical heating and energy non-conservation, while conserving the charge density (locally) and total momentum exactly.
We find that even an egregious violation of the condition in equation~\eqref{eq:growth-condition} still allows for significant growth, but at slower rates.
The effect of the instability (during the linear evolution) on the beam energy loss is similar to that in the homogeneous cases.
The reason is that, in presence of background plasma inhomogeneities, the system stays for longer times in the linear regime, saturating at similar levels.

Our one dimensional simulations present an idealized experiment to study the effect of the background plasma inhomogeneity on the growing wavemodes in the case of blazar driven pair beams for the following reasons.
First, only longitudinal wavemodes --- whose growth is largely insensitive to the details of the momentum distribution of the pair-beams~\citep{Bret-2010-POP} --- are included in our simulations.
Therefore, simulations that initially resolve the spectral support of the instability will be able to resolve the instabilities during the physical evolution that typically results in increasing the width of the momentum distribution of the pair-beams~\footnote{
In higher dimensions, the fastest growing wavemodes are oblique wavemodes, whose spectral width are very sensitive to the details of the beam momentum distribution~\citep{Bret-2010-POP,Timofeev+2009}, thus {\it correctly} capturing the instabilities during the physical evolution is a challenging computational problem.
}.
Second, in the linear regime, the effect of the background plasma inhomogeneity is fully decoupled from the momentum distribution of the background and pair beams.
Thus, the effect of inhomogeneity on the growth of the other wavemodes (oblique and Weibel modes) is expected to be very similar to that on the longitudinal wavemodes.
That is, for extreme beam parameters, {\it correctly} simulating the physical evolution is computationally tractable only for the longitudinal wavemodes, and since the effect of inhomogeneity  is similar for all unstable wavemodes, our results is expected to hold when the Oblique and Weibel wavemodes are correctly captured in simulations as well.
However, we leave explicit demonstration of this point to future work.

Here, we focus on the case of relativistic dilute pair-beams relevant for blazar-induced pair-beams~\citep{blazarI}.
However, we note that numerical studies of the inhomogeneity effects\footnote{These include both simulations and numerical solutions of the Zakharov equations (approximate nonlinear evolution equations).} on the electrostatic wavemodes (1D) of nonrelativistic beam plasma instabilities have been previously performed in the context of solar wind plasmas~\citep[see, e.g.,][]{Inhom01,Inhom02,Inhom03,Inhom04,Krafft-2013,Krafft-2015,Krafft-2017}.

This paper is organized as follows:
Section~\ref{sec:problem} defines the problems with the linear perturbation analyses for beam-plasma system in presence of inhomogeneity.
In Section~\ref{sec:linear-rates}, we discuss the condition for the validity of linear growth rates in
inhomogeneous plasmas and derive a general condition for growth of longitudinal wavemodes.
In Section~\ref{sec:sims}, we present the setup for simulations that violate this condition and discuss their numerical convergence.
In Section~\ref{sec:results} we present our simulation results in the linear and saturated, non-linear regimes.
We conclude  in Section~\ref{sec:conclusions}.

\section{Non-uniform background plasmas: defining the problem}
\label{sec:problem}

For a beam-plasma system with a fixed neutralizing background, we denote the phase space distribution functions of beam electrons/positrons by $f^{\pm}$ and for background electrons by $g$. The linearized (first-order) Vlasov-Maxwell equations, which describe the longitudinal evolution of a linear perturbation, are given by
\begin{align}
\label{eq:vm-system}
&
\partial_t f^{\pm}_1(x,t,u) + v \partial_x f^{\pm}_1(x,t,u) \pm \frac{e}{m_e} E_1(x,t) \partial_u f^{\pm}_0(u) = 0,
\\
&
\partial_t g_1(x,t,u) + v \partial_x g_1(x,t,u) - \frac{e}{m_e} E_1(x,t) \partial_u g_0(x,u) = 0,
\\
&
\partial_x E_1(x,t) = \frac{ e }{\epsilon_0} \int \left[ f^{+}_1(x,t,u) - f^{-}_1(x,t,u) - g_1(x,t,u) \right] du,
\end{align}

where, $e$ and $m_e$ are the elementary charge and mass of electrons, $v$ is the velocity in phase space, $u = \gamma v$ with $\gamma = 1/\sqrt{1-v^2}$,
$f^{\pm}_{0}$ and $f^{\pm}_{1}$ are the equilibrium and the first order perturbation of the phase space distribution function of pair-beam  plasma particles, respectively, $g_{0}$ and $g_{1}$ are the equilibrium and the first order perturbation of the phase space distribution function of background electron plasma, respectively, and $E_1$ is the first order perturbation in the electric field.

Due to the inhomogeneously distributed background electrons, the equilibrium distribution function $g$ depends on the position $x$.
To solve these equations as an initial value problem (using the Landau procedure), one takes the Fourier/Laplace transform for $x/t$ and  assumes initial perturbations for the pair-beam plasmas $f^{\pm}_{\rm ini} = f_1^{\pm}(x,u,t=0)$ and for the electron background plasma $g_{\rm ini} = g_1(x,u,t=0)$~\citep{Landau:1946,Nicholson1983,Boyd}).
We get the following equations:
\begin{align}
i k  E_1(k, \omega) &= \frac{ e }{\epsilon_0} \int \left[ f^{+}_1  - f^{-}_1  - g_1  \right] du,
\\
\label{eq:Bmf}
f^{\pm}_{\rm ini}(k,u) &=
- i (\omega - k v ) f^{\pm}_1(k,\omega,u)   \pm \frac{e \partial_u  f^{\pm}_0(u) }{m_e} E_1(k,\omega) ,
\\
\label{eq:Bgg}
g_{\rm ini}(k,u) &=
- i (\omega - k v ) g_1(k,\omega,u)
\nonumber \\
& ~~~
- \frac{e}{m_e} \int dk^{'} E_1(k-k^{'}, \omega)  \partial _u  g_0(k^{'},u),
\end{align}
where, $f^{\pm}_{\rm ini}(k,u)$ and $g_{\rm ini}(k,u) $ are the Fourier transform of $f^{\pm}_{\rm ini}(x,u)$ and $g_{\rm ini}(x,u) $, respectively.
Therefore,
\begin{align}
\label{eq:Eevol}
&
\left[
k
+
\dfrac{ e^2}{m_e \epsilon_0 }
\int \frac{du }{ \omega - k v }
\partial _u (f^{+}_0 + f^{-}_0)
\right]
E_1 (k,\omega)
\nonumber \\
&
~~~~~~~~~~~~~~
+
\dfrac{e^2}{m_e \epsilon_0 }
\iint
dk^{'}
du 
\frac{ ~ \partial _u  g_0(k^{'},u)
}{ \omega - k v }
E_1(k-k^{'}, \omega)
\nonumber \\
&~~~~~~~~~~~~~~~~~~~~~~~~~~~~~~~~~~~~~~
=\dfrac{e}{ \epsilon_0 }
\int \frac{  du } {\omega - k v }
\left[
f^{+}_{\rm ini}
-
f^{-}_{\rm ini}
-
g_{\rm ini}
\right].
\end{align}
The convolution in Equation~\eqref{eq:Eevol} implies coupling between all Fourier modes of $E_1(k,\omega)$ with all Fourier modes in the background plasma inhomogeneity, i.e., a nonlinear coupling of the modes to the structure in the background.
This is a direct consequence of the fact that, in general, the normal wavemodes of an inhomogeneous plasma are not Fourier modes.
In other words, the Fourier modes do not describe linearly independent solutions to the linearized equations.

\section{Applicability of homogeneous plasma linear growth rates in presence of inhomogeneity}
\label{sec:linear-rates}

First, we follow the empirical discussion in \citet{Breizman+Inhomo01} and \citet{Miniati-Elyiv-2013} and quantify the timescales that determine whether the homogeneous plasma linear growth rates are applicable.
Then, we derive the applicability condition for uniform plasma growth rates of longitudinal unstable wavemodes in presence of inhomogeneity.

\subsection{General discussion}

To simplify, we consider only inhomogeneity along the pair-beam direction.
The timescale on which inhomogeneity affects the growing wavemode can be found using the geometric optics approximation:
\begin{eqnarray}
\frac{ d k }{ dt } = \frac{ d \omega }{ dx } \sim \frac{ \omega_g }{ 2 \lambda^{\rm inh}_{\parallel} },
\end{eqnarray}
where $\lambda^{\rm inh}_{\parallel}$ is the inhomogeneity scale length in the pair-beams direction and $\omega_g$ is the background plasma frequency at mean density.
Therefore, if the width of  unstable oblique wavemodes along the beam direction is $\Delta k_{\parallel}$, then the inhomogeneity timescale is given by
\begin{eqnarray}
\tau_{\rm inh}
\equiv
\frac{\Delta k_{\parallel}} { | d k / dt|}
\sim
\frac{ \Delta k_{\parallel} 2 \lambda^{\rm inh}_{\parallel} }{ \omega_g }.
\end{eqnarray}

The timescale of oblique wavemodes linear growth is given by
\begin{eqnarray}
\tau_{\rm obl}
\equiv
\frac{ \Lambda^{^{\rm obl}} } { \Gamma^{^{\rm obl}}  },
\end{eqnarray}
where $\Lambda^{^{\rm obl}}$ is the number of $e$-foldings  for oblique wavemodes before the nonlinear effects become important and $\Gamma^{^{\rm obl}} $ is the maximum growth rate of oblique wavemodes.
Therefore, the condition for the validity of linear homogeneous growth rate in the presence of background inhomogeneity, $\tau_{\rm inh} \gg \tau_{\rm obl}$, corresponds to
\begin{eqnarray}
\label{eq:inhomo-cond}
 \lambda^{\rm inh}_{\parallel}  \gg  \lambda_{\rm min} \equiv \frac{ \Lambda^{^{\rm obl}}  } { 2} \frac{\omega_g } { \Gamma^{^{\rm obl}}  } \frac{1}{  \Delta k_{\parallel} }.
\end{eqnarray}

Previously, \citet{Miniati-Elyiv-2013} argued that the violation of equation (\ref{eq:inhomo-cond}) results in a severe suppression of the linear growth rate.
We now test this suppression using a series of 1D numerical simulations in which the condition is violated.
We find that the maximum growth rate is slower than predicted for the homogeneous case but the unstable wavemodes still grow and the pair beam energy loss (in the linear growth) is approximately the same as that of the homogeneous plasma case
\footnote{
We note  that the spectral width for oblique unstable wavemodes, $\Delta k_{\parallel}$, used in \citep{Miniati-Elyiv-2013,Breizman+Inhomo01} reduces to
$\Delta k_{\parallel} = 0$ in the cold-limit.
However, when the cold-limit dispersion relation is solved a finite spectral width of wavemodes growing with rates comparable to the maximum growth rates exits~\citep[see, e.g.,][]{Bret-2010-PRE}.
}.

\subsection{Longitudinal unstable wavemodes}

Here, we  derive a condition for the validity of the electrostatic growth rates (in the cold limit) using the correct spectral width found by~\citet{resolution-paper}.
For a uniform background, the dispersion relation for Langmuir (longitudinal) waves, in the cold-limit, is given by ~\citep{fainberg+1969}
\begin{align}
\label{eq:dis_cold}
1-  \dfrac{1}{\hat{\omega }^2}
- \dfrac{\alpha /\gamma _b^3 }{  (\hat{\omega }-\hat{k} )^2}
=
0,
\end{align}
where, $\hat{\omega} = \omega / \omega_g$, 
$\hat{k} = k v_b / \omega_g$, $v_b$ is the beam velocity, $\alpha = n_b/n_g$, $\omega_g = \sqrt{n_g e^2/m_e \epsilon_0}$ is the background plasma frequency and $n_g$ and $n_b$ are the number densities of background and beam plasma, respectively.
The dispersion relation in the linear regime, Equation~\eqref{eq:dis_cold}, implies instability (exponential growth) for  all wavemodes with wavelengths \citep{resolution-paper}
\begin{align}
\lambda \geq \frac{  2 \pi v_b }{\omega_g} \left(1 + \frac{ \alpha^{1/2} }{\gamma_b} \right)^{-3/2}
~~
\Rightarrow
~~
k  \leq  \left(1 + \frac{ \alpha^{1/2} }{\gamma_b} \right)^{3/2} \frac{\omega_g }{v_b}.
\end{align}

The spectral width, $\Delta k_{\parallel}$, of these unstable wavemodes along the beam direction is given by~\citep[using Equation 10 of][]{resolution-paper}
\begin{eqnarray}
\Delta k_{\parallel}   \approx \Delta k_{1/2} = \frac{2 \pi }{1.15008} \left( \frac{\alpha}{\gamma_b^3} \right)^{1/3} \omega_g/c,
\end{eqnarray}
where, $\Delta k_{1/2}$ is the full-width-half-max of unstable wavemodes, i.e.,
it is the k-space width of unstable wavemodes that grow with rates $ \geq 0.5$ of the maximum growth rate.
The maximum growth rate for these wavemodes is given by~\citep{Bret2010}
\begin{eqnarray}
\label{eq:long_rate}
\Gamma^{^{\rm L}}   \sim \frac{ \sqrt{3} }{ 2^{4/3} } \left( \frac{\alpha}{\gamma_b^3} \right)^{1/3} \omega_g .
\end{eqnarray}
Therefore, for the longitudinal unstable wavemodes
\begin{align}
\label{eq:lmin-Longi}
\lambda_{\rm min}
&=
\frac{ \Lambda^{^{\rm L}}  } { 2} \frac{  2^{4/3} } { \sqrt{3}  } \frac{1.15008}{  2 \pi }  \left( \frac{\alpha}{\gamma_b^3} \right)^{-2/3} \frac{c}{\omega_g}
\nonumber
\\
&=
1.33
\left( \frac{ ~ \Lambda^{^{\rm L}} }{10} \right)
\left(		 \frac{\alpha}{\gamma_b^3}	 \right)^{-2/3}
~ c / \omega_g .
\end{align}

\section{Numerical Simulations -- setup and convergence}

\label{sec:sims}

Here, we present a number of 1D simulations with inhomogeneity that strongly violate the condition, $\lambda^{\rm inh}_{\parallel} \gg \lambda_{\rm min}$, with $\lambda_{\rm min}$ given in Equation~\eqref{eq:lmin-Longi}.
The growth of unstable wavemodes and energy loss from the pair-beams are contrasted with the corresponding results of a uniform background plasma simulation.
While we find that the growth rates are reduced modestly in presence of background plasma inhomogeneity,  the pair-beams energy loss is found to be similar in all simulations.

\begin{figure}
\center
\includegraphics[width=8.6cm]{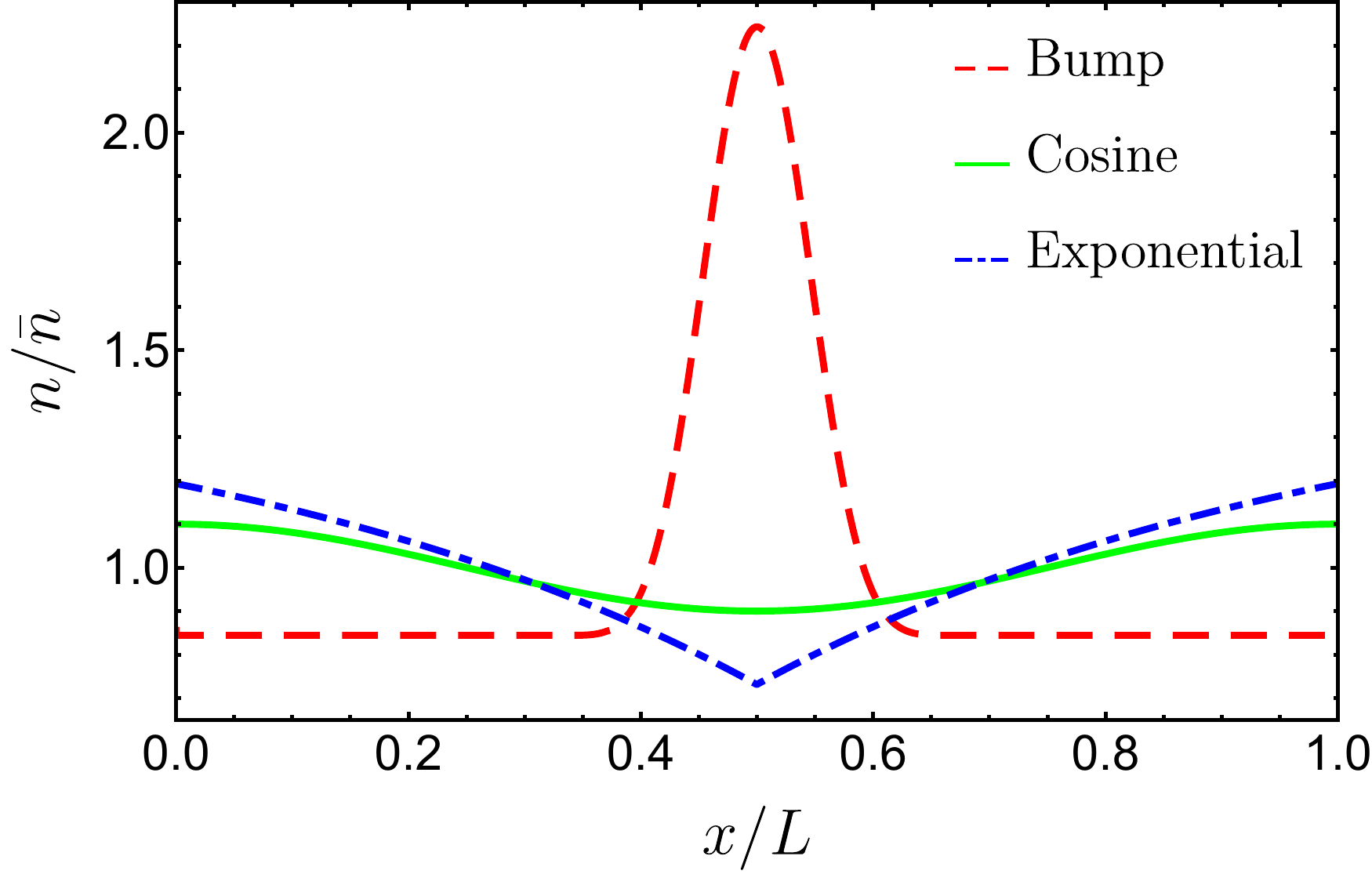}
\caption{
Initial background plasma number density (electrons and fixed ions) for the various inhomogeneous simulations.
All simulations are 1D with a computational domain of length $L$.
For each, the number density $n$ is normalized by the average number density $\bar{n}$.
\label{fig:IniDen}
}
\end{figure}

\begin{deluxetable*}{
cccccc
}
\tablewidth{18cm}
\tablecaption{
Electrostatic Beam-Plasma instability simulations with $\alpha = 0.002$, $\gamma_b=100$.
\label{table:inhsim-para}
}
\tablehead{
Simulation             				&
$L_c$~\tablenotemark{a}					&
$\Gamma^{\rm sim} / \Gamma^{\rm th}_{\rm max} $\tablenotemark{b} &
$N_{\rm pc} $  \tablenotemark{c} &
$\lambda^{\rm inh}_{\parallel} $  \tablenotemark{d} &
$E_r/E_{b,0}$ \tablenotemark{e}
}
\startdata
\rule{0pt}{8pt}
``Uniform'' &  $263.9$ & $ 0.883$ & $2706.4$ 		&	$\infty$
& 0.0017 \%
\\
\rule{0pt}{8pt}
``Bump'' & $861$   &	$ 0.75 $ & $3491.3$	&	$\sim 80 ~c/\omega_p$
& 0.0013 \%\\
\rule{0pt}{8pt}
``Cosine'' & $128$ 	& 	$ 0.20$ & $3914.0 $		& 	$\sim 200 ~c/\omega_p$
& 0.0012 \%
\\
\rule{0pt}{8pt}
``Exponential'' &  $125$ & $ 0.10$ & $4008.0$ 		&	 $\sim 62.5 ~c/\omega_p$
&  0.0005 \%
\enddata
\tablenotetext{a}{The box size, $L$, in units of skin depth, i.e., $ L_c = L ~ \omega_p / c$, where, $\omega_p$ is the plasma frequency associated with all plasma particles: beam and background particles.}
\tablenotetext{b}{The maximum growth rate found in simulations $\Gamma^{\rm sim}$ normalized to the maximum growth rate predicted theoretically for a uniform plasma, $\Gamma^{\rm th}_{\rm max}
= 8.647 \times 10^{-4} ~ \omega_p$, found by solving the dispersion relation in Equation~\eqref{eq:dis_cold}.}
\tablenotetext{c}{Total number of macro-particles (background electrons and beam electrons and positrons) divided by the total number of computational cells.}
\tablenotetext{d}{The inhomogeneity scale length; the scale length on which the background plasma number density changes significantly. To obtain the degree of violation for the condition of the validity of homogeneous growth rates ($\lambda^{\rm inh}_{\parallel} \ll \lambda_{\min}$), this should be compared to $\lambda_{\rm min} \sim  8.38 \times 10^5	 ~c/\omega_p  $, i.e., this condition is violated by about three order of magnitudes in all non-uniform  simulations.
}
\tablenotetext{e}{ Maximum energy error in simulations normalized to the initial energy of the pair-beam particles.
}
\end{deluxetable*}

\subsection{Simulation setup}

For our numerical simulations, we use SHARP-1D~\citep{sharp} with fifth order interpolation, $W^5$, to improve the conservation of energy in simulations while conserving the total momentum exactly.
Using SHARP with $W^5$ is essential to avoid the excessive numerical heating typical in most available PIC codes. Importantly it eliminates numerical heating
for long-time simulations \citep[millions of $\omega_p^{-1}$, see][for illustration]{sharp}.

In all simulations, we resolve the plasma skin depth, $c/\omega_p$, by 20 cells, i.e., the cell size is $\Delta x = 0.05~ c/\omega_p $, and use a time step that satisfies the Courant-Friedrichs-Lewy (CFL) stability condition; we used $c \Delta t / \Delta x = 0.4$.
The momentum distributions of the beam and background plasmas (in their individual comoving frames) are initialized using a thermal distribution with normalized temperatures $\theta_g = \theta_b = 4 \times 10^{-3}$, where, $\theta = k_B T / m_e c^2$.
In all cases periodic boundary conditions are applied, and other simulation parameters are laid out in Table~\ref{table:inhsim-para}.

\begin{figure}
\center
\includegraphics[width=8.6cm]{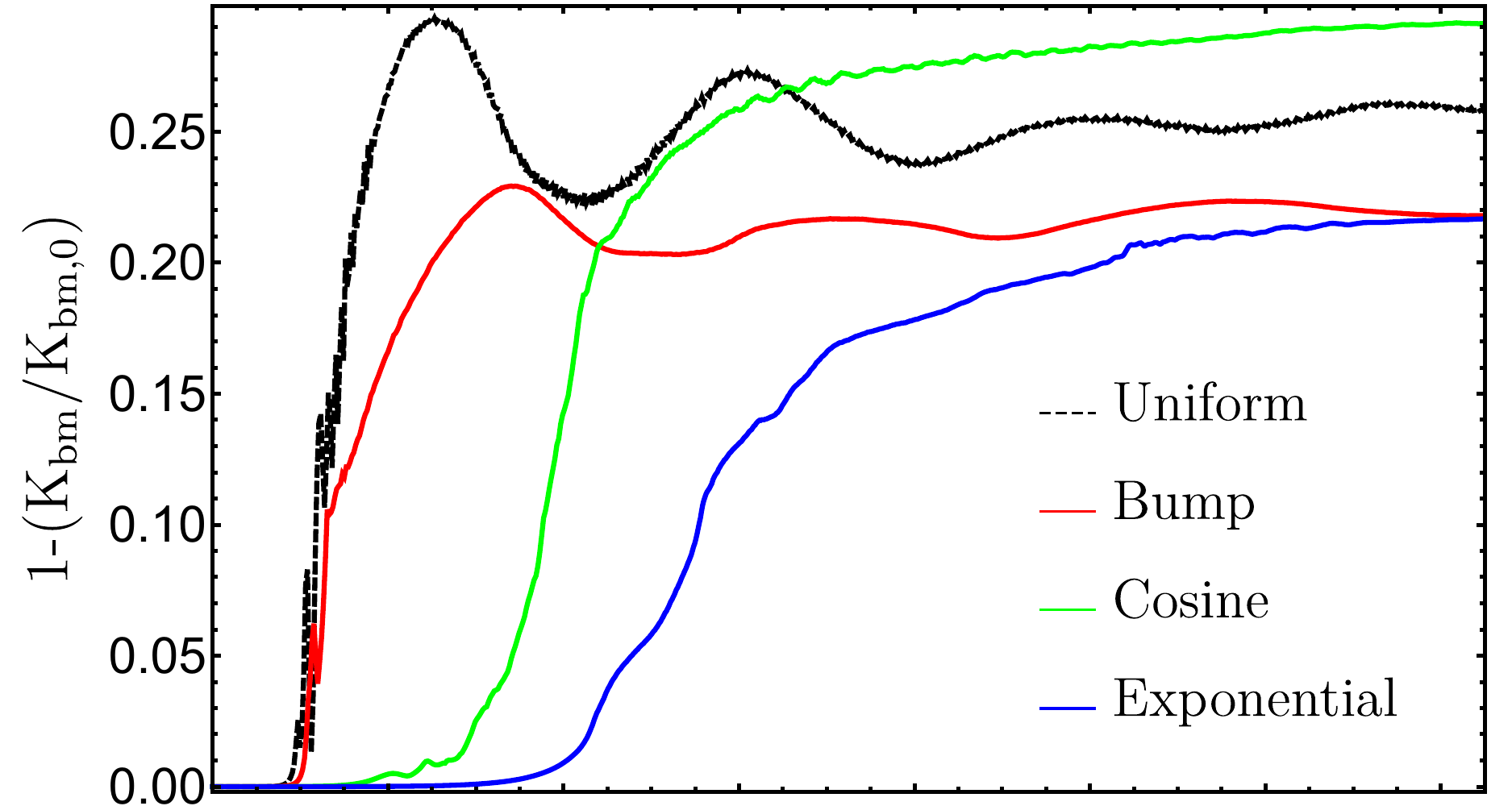}
\\
\includegraphics[width=8.6cm]{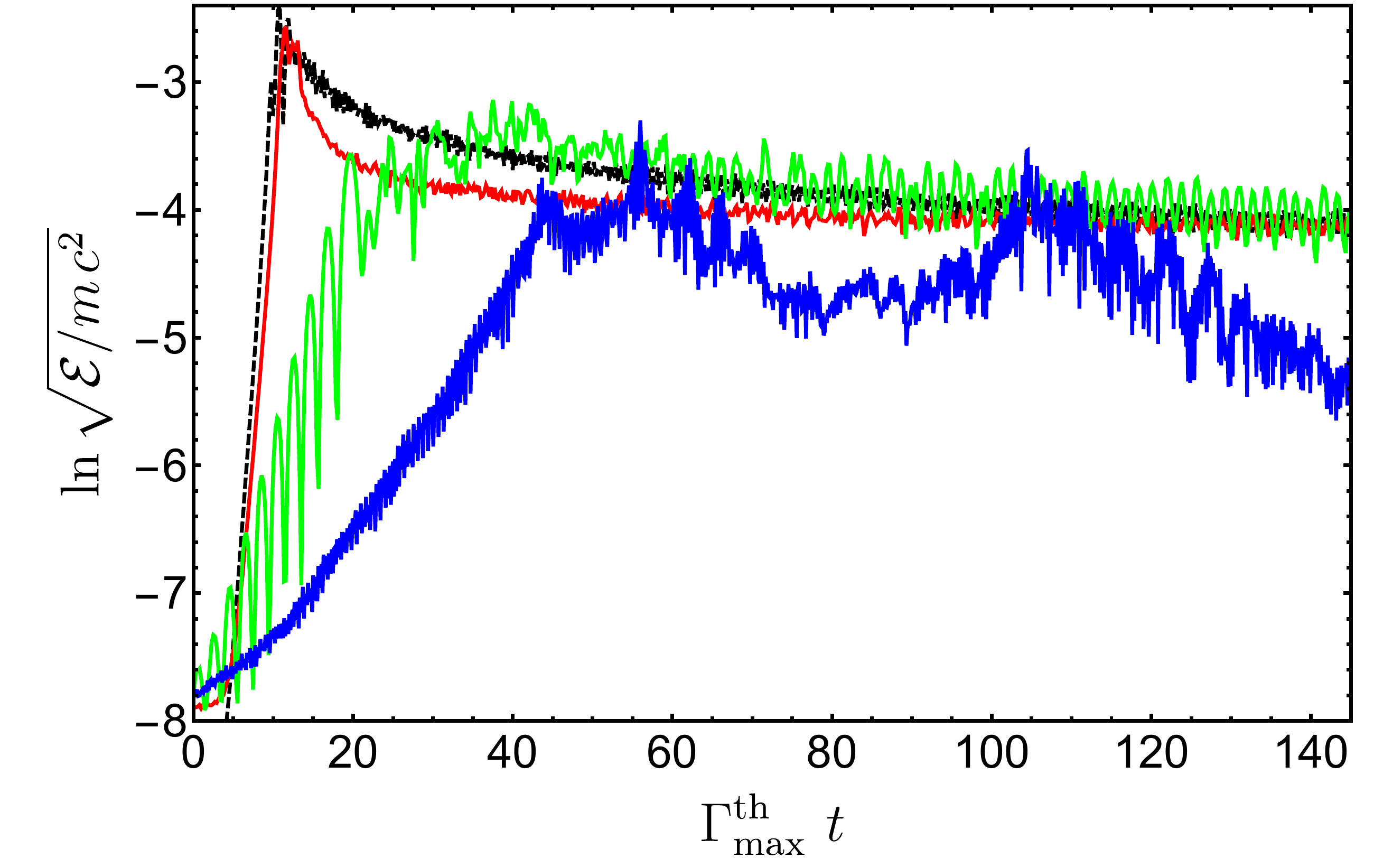}
\caption{
Top: evolution of the fractional energy loss of beam particles in different simulations with inhomogeneous and homogeneous background plasmas.
Bottom: evolution of the total potential energy per computational particle, $\mathcal{E}$, normalized to the rest mass energy of a computational particle,  $m \hspace{.05cm} c^2$.
Since the growth in all simulations starts from the Poisson noise, the times are shifted in different simulations
(by a maximum of $\Gamma_{\rm max}^{\rm th} t=7$, depending on resolution)
to allow a direct comparison of the exponential growth rates of the potential energy. 
\label{fig:SimRes}
}
\end{figure}

We assume a background plasma (electrons and {\it immobile} ions) that is spatially inhomogeneous, but charge neutral, in all simulations.
We also assume a spatially uniform pair-beam plasma (electrons and positrons) that moves with a Lorentz factor of $\gamma_b = 100$, and a beam to background ratio of $\alpha=n_b/n_g=0.002$, where $n_b$ and $n_g$ are the number density of beam particles and background electrons, respectively.\footnote{Note, here $n_b$ is the number density of all beam particles (both electrons and positrons).}. 
As a result, the pair beam is highly relativistic but energetically subdominant, similar to those anticipated in the IGM.
For all simulations,
\begin{align}
\lambda_{\rm min} \sim  8.38 \times 10^5	 ~c/\omega_p,
\end{align}
where, we set\footnote{Typically, it is assumed that $\Lambda^{\rm L} \sim 30$~\citep{Huba2013,Miniati-Elyiv-2013}. 
However, we set $\Lambda^{\rm L} = 10$, since we observe from the second panel of Figure~\ref{fig:SimRes} that it is of order 10.
Setting it to higher values implies higher values for $\lambda_{\rm min}$, and, thus, a stronger violation of the condition in Equation~\eqref{eq:inhomo-cond} by the same factor.} $\Lambda^{\rm L} = 10$.
Since, in all inhomogeneous simulations, the inhomogeneity scale length $\lambda^{\rm inh}_{\parallel} \lesssim 200~c/\omega_p$, the condition in Equation~\eqref{eq:growth-condition} is violated by more than three orders of magnitude in all cases.

We consider three cases which fit within two classes of background inhomogeneities, distinguished by the extent of the background variations.
The first of these is the ``Bump'' simulation; a simulation with a central Gaussian bump in the background density.
The periodic simulation domain is divided into three parts, in the first and last part the number density is uniform, while in the middle part, the number density follows a Gaussian distribution with standard deviation, $\sigma = (L/20)$; red curve in Figure~\ref{fig:IniDen}).

Within the second class, we present two simulations with a continuously varying background density.
In the first, the background plasma number density is varying as a cosine across the box (with amplitude, $A=0.1$); green curve in Figure~\ref{fig:IniDen}. We call this the ``Cosine'' simulation.
In the second, the variation is much faster; the number density at the edges of the computational domain is about $1.2$ of the average number density, $\bar{n}$, and drops exponentially fast until it reaches
$0.73~\bar{n}$ in the middle of the computational domain and then increases exponentially fast afterwards to reach $1.2~\bar{n}$ at the other edge of the computational domain (see blue curve in Figure~\ref{fig:IniDen}).
We call this the ``Exponential'' simulation. More preciously, the number density in the ``Exponential'' simulation is given by
\begin{eqnarray}
\frac{ n(x/L) }{\bar{n} }
& = &
\frac{2 e}{1+e}-\frac{e^{1-2 | x/L ~ - ~0.5 | }}{1+e}.
\end{eqnarray}

Since the theoretically expected maximum growth rate
is $\Gamma_{\rm max}^{\rm th} = 8.647 \times 10^{-4} \omega_p$, during one growth time, beam particles travelling with $v_b \sim c$ will travel a distance $\sim 1156 ~ c/\omega $. That is, during one growth time, the beam particles will travel distances larger than the box size in all simulations.
Since the growth rates in nonuniform simulations is lower than $\Gamma_{\rm max}^{\rm th}$ (see Table~\ref{table:inhsim-para}), the distance travelled by the beam, during one growth time in the simulation is even larger than $1156 ~ c/\omega_p$.

\begin{figure*}
\center
\includegraphics[width=14cm]{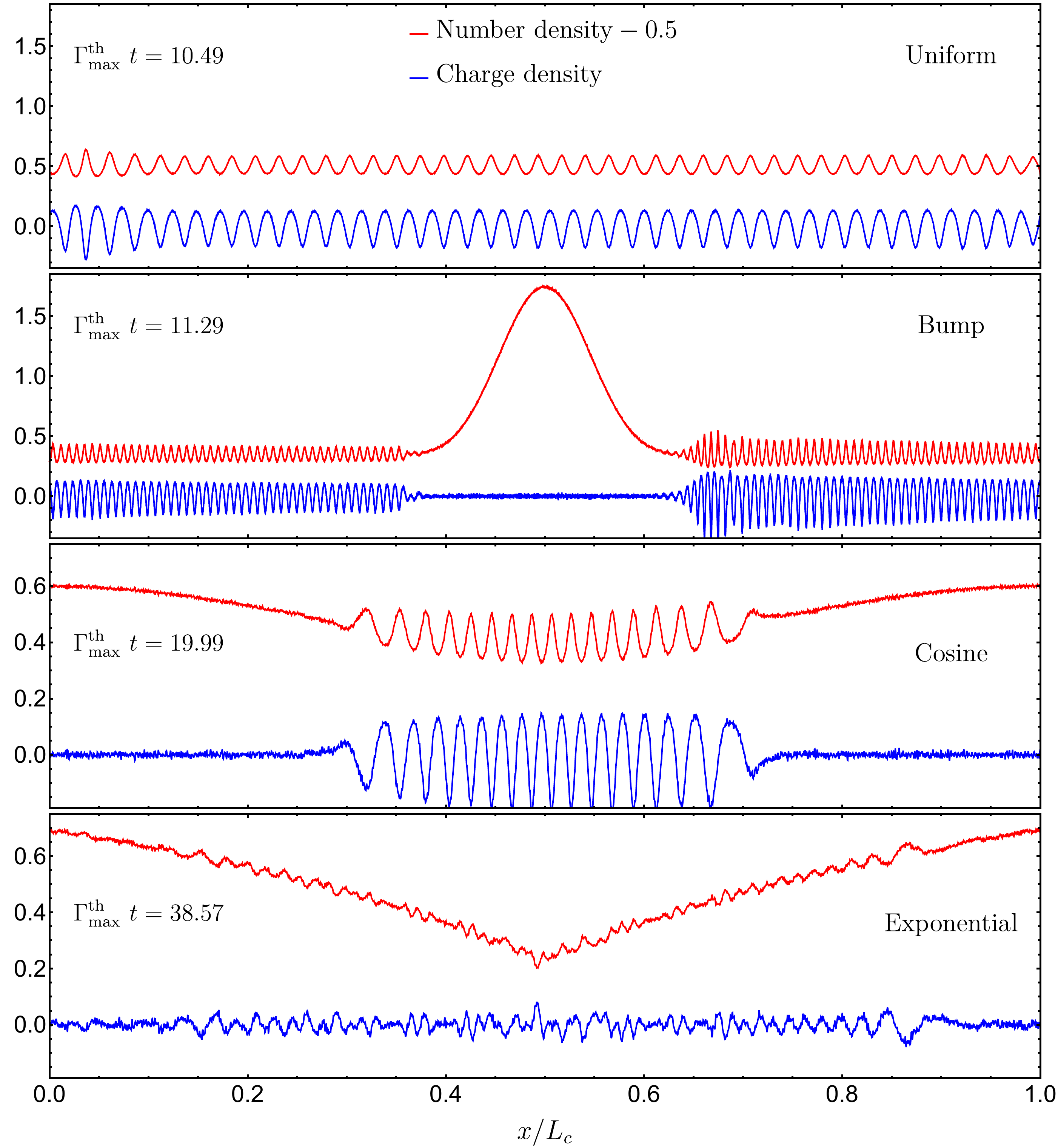}
\caption{
Instantaneous charge density (red) and number density (blue) close to the end of the linear evolution for the different inhomogeneous and homogeneous background plasma simulations.
In all cases, unstable wavemodes are excited despite variations in the background density.
\label{fig:PlWaves}
}
\end{figure*}

\subsection{Simulation convergence and performance}

Convergence of numerical simulations is an essential way to avoid confusing the evolution of different numerical errors with physical evolution.
In~\citet{sharp}, we demonstrated that the typical method of checking the convergence in PIC simulations is misleading: increasing the number of particles per cell ($N_{\rm pc}$) and decreasing the cell sizes ($\Delta x$) independently was shown to lead to a plateau in the numerical errors.
Thus changing these parameters independently does not imply convergence as typically claimed.
The correct convergence was shown to be only possible when both of these resolution criteria are improved simultaneously.

We follow this approach to check the convergence for all inhomogeneous simulations presented here, and present the results from the highest resolved simulations.
For all inhomogeneous simulations presented here, the beam energy evolution and the grid potential energy evolution are very similar to that in simulations with resolution lower by a factor of $2$, i.e., with $0.5 N_{\rm pc}$ and $2 \Delta x$.

For uniform simulations, the uniformity coupled with the periodicity on the physical domain imply a minimal spectral width within which wavemodes are not resolved in simulations.
This means that there is another resolution criteria (box size $L$) that should be also improved (independently or simultaneously with other resolution criteria) in order to resolve the narrow spectral width of the unstable wavemodes of relativistic and dilute pair-beams instabilities~\citep{resolution-paper}.
Thus, for uniform simulations, we perform simulations with resolution increased by factors of $2$ and $4$, i.e., increase $L$ and $N_{\rm pc}$ and decrease $\Delta x$ by such a factor simultaneously.
In all simulations we obtain very similar pair-beam energy evolution and grid potential energy evolution, we present results of the lower resolution uniform simulation here to facilitate comparisons with inhomogeneous simulations\footnote{A uniform simulation with a box size smaller by a factor of $2$ compared to the uniform simulation presented here results in very different energy evolution and lower energy saturation level ($\sim 14\%$) and also results in a slower growth of the potential energy on the simulation grid.}.

As pointed out above, the use of higher order interpolation functions greatly reduce the numerical heating typical in PIC simulations. The maximum energy errors (normalized to the initial beam energy) in our simulations are always below 0.002~\%, i.e., the energy error is less than $2 \times 10^{-5}$ of the initial beam energy, and less than 0.1\% of the background thermal energy (see Table~\ref{table:inhsim-para} for details).

\begin{figure*}
\centering
\begin{tabular}{ll}
\includegraphics[width=0.49\textwidth]{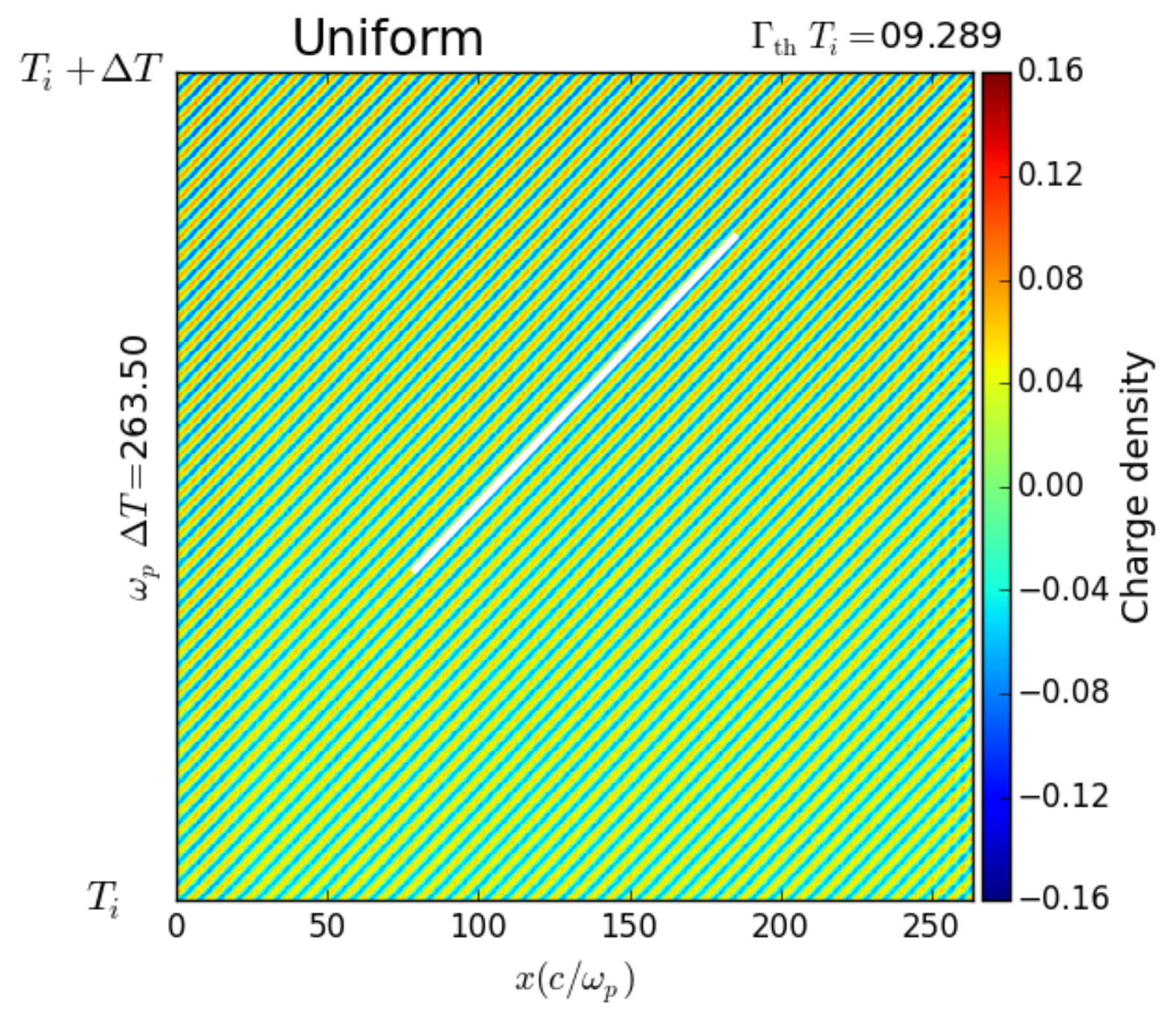}
&
\includegraphics[width=0.49\textwidth]{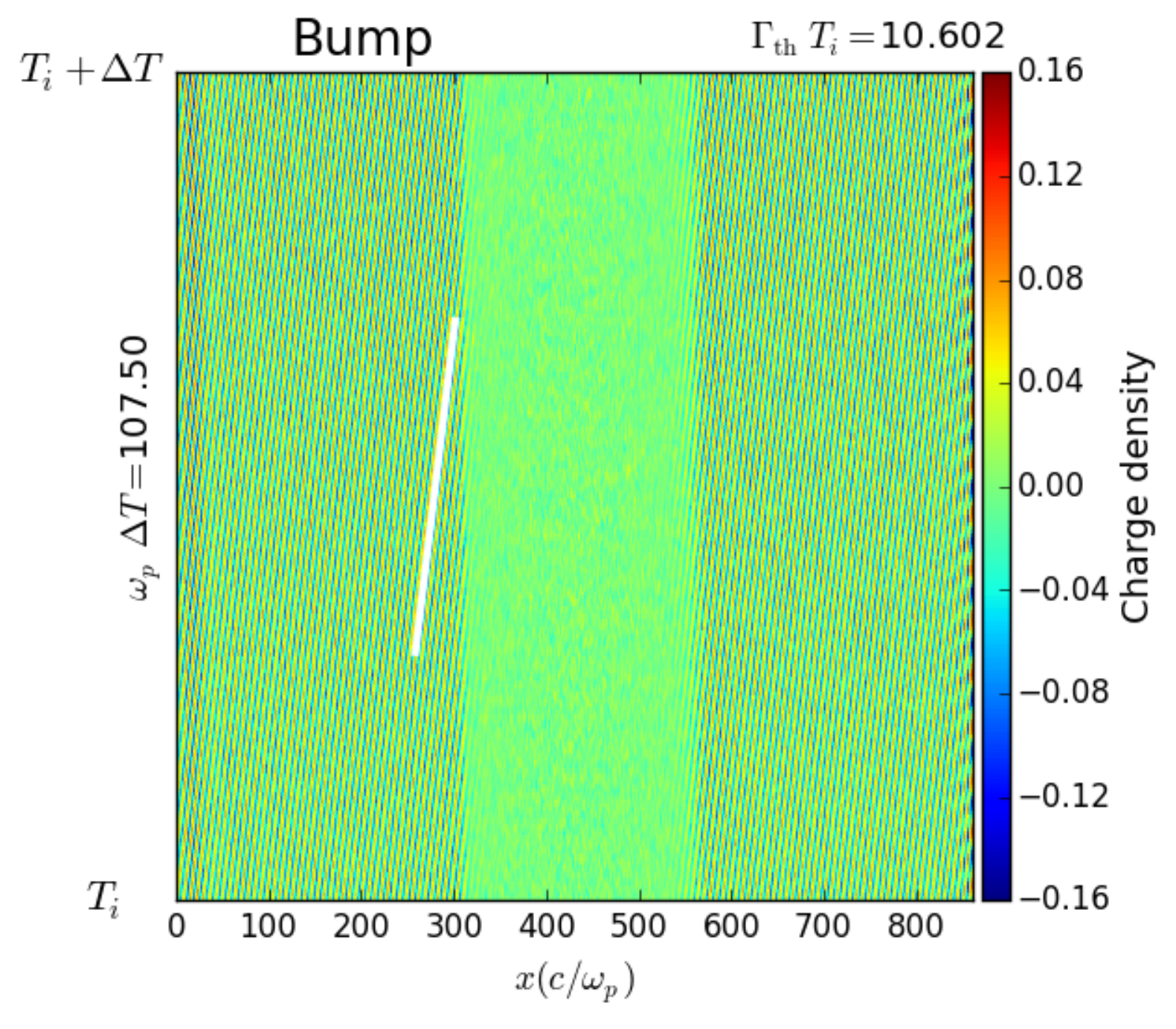}
\\
\includegraphics[width=0.49\textwidth]{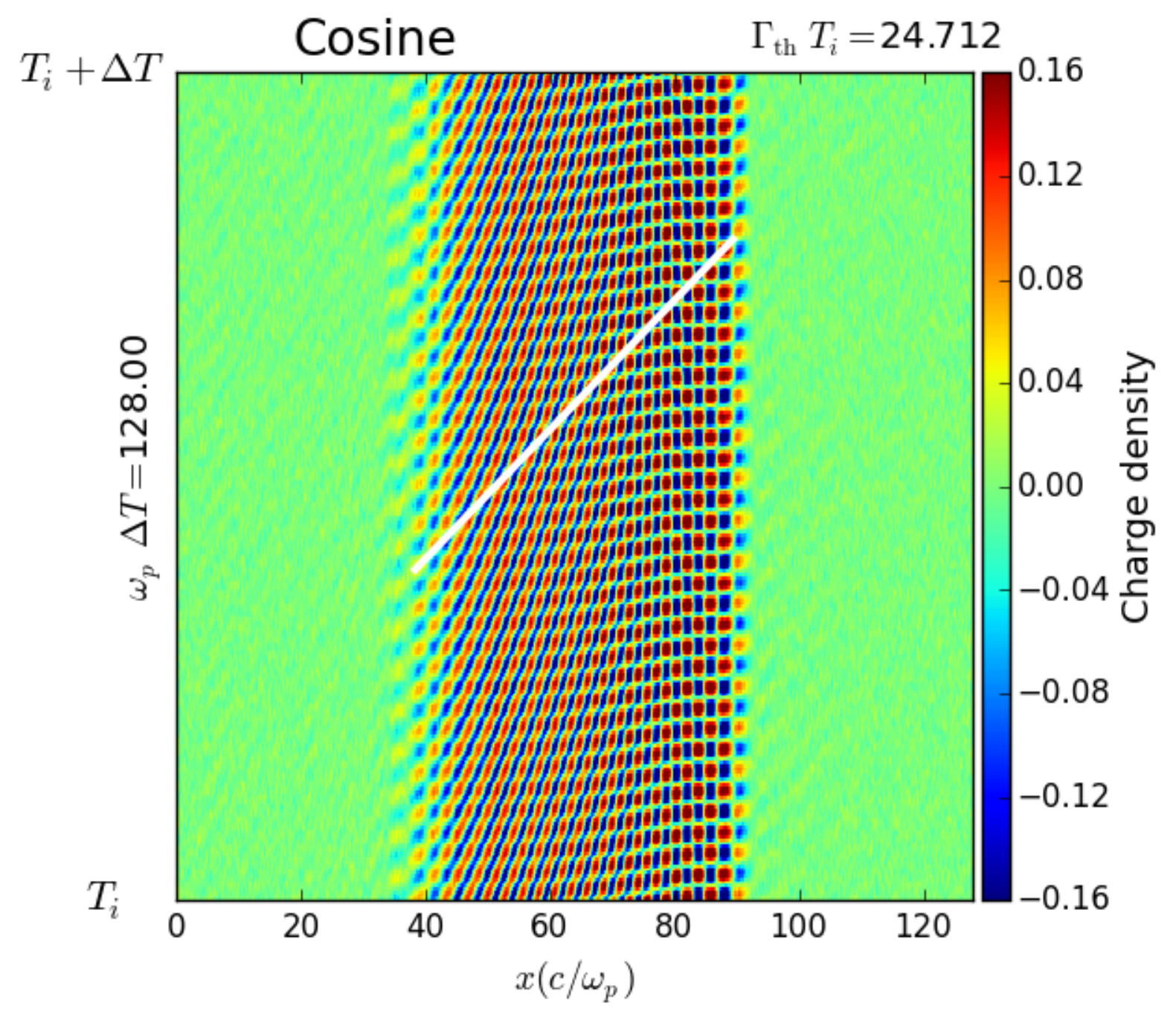}
&
\includegraphics[width=0.49\textwidth]{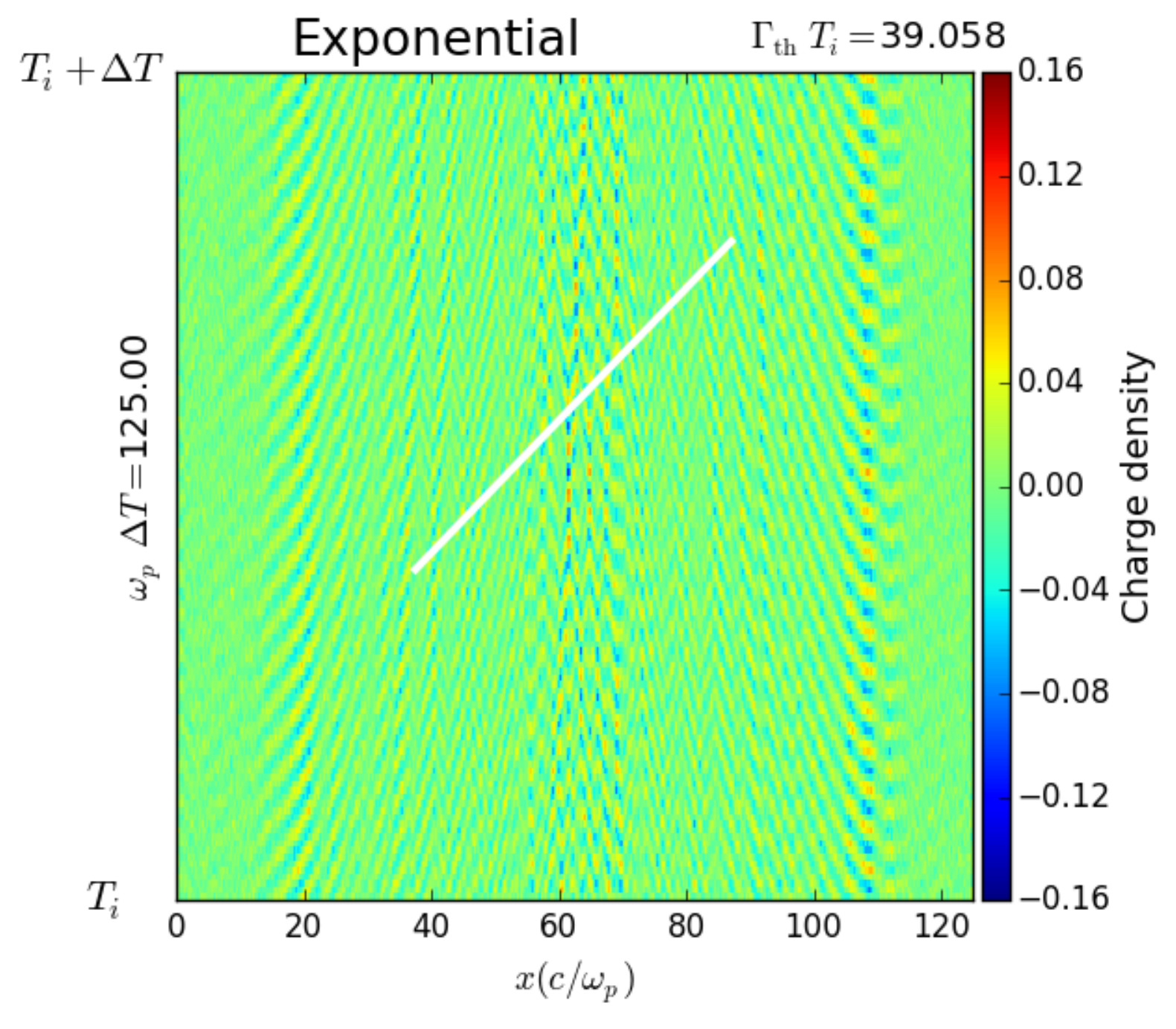}
\\
\end{tabular}
\caption{
Growth and evolution of the charge density for different simulations close to the end of the linear regime evolution.
The charge density is color coded as a function of position in the simulation box and and over a time period $T_i < t < T_i+\Delta T$ as indicated in the labels in each panel.
In all panels, wavemodes that travel along white lines are travelling with the speed of light along the direction of the beam ($\sim v_b$). 
Checkerboard patterns indicate standing waves, and thus the presence of both forward- and backward-propagating wavemodes.
\label{fig:denxt1}
}
\end{figure*}

 \section{Simulation Results}
\label{sec:results}

\subsection{Linear regime evolution}

In Figure~\ref{fig:SimRes}, we show the evolution of the fractional beam energy loss (top) and the electric potential energy (bottom).
The potential energies grow at smaller rates of about $0.75$, $0.2$, and $0.1$ of the maximum growth rate for uniform plasmas for ``Bump'', ``Cosine'' and ``Exponential'' simulations, respectively.
Figure~\ref{fig:PlWaves} shows the growth of unstable plasma wavemodes even when the condition is severely violated.

In all simulations, the level at which the beam energy loss stops occurring with rates comparable to the linear growth rates is the same, i.e., about $20-26$\% energy loss.
This is similar to the level of saturation we obtained from a uniform background simulation with the same beam parameters.
The growth rate of the uniform simulation is 0.883 of the maximum growth rate predicted from theory.
This is in perfect agreement with the maximum growth rate for wavemodes allowed to grow in such a simulation box with periodic boundary conditions, see~\citet{resolution-paper} for more details.

For the ``Bump" simulation, the inhomogeneity scale length is between that of ``Cosine" and ``Exponential" simulations.
However, since 2/3 of the box in the ``Bump" simulation is uniform, the effective growth rate is higher than in the cases where the background number density is varying throughout the simulation domain.
As seen in the second panel of Figure~\ref{fig:PlWaves}, displaying the ``Bump" simulation, the growth of plasma waves is localized in the uniform regions despite the fact that the group velocity of the growing wavemodes is equal to the beam velocity ($\sim c$).
This implies that the initially excited forward-propagating wavemodes reflect at high density regions.

Figure~\ref{fig:denxt1} shows further evidence that growing modes reflect as they propagate toward high-density regions, which shows a representative period for the charge density during the linear regime  evolution for each simulation.
For the ``Uniform" simulations, the fastest growing wavemode is a propagating wave with group velocity equal to the beam velocity ($\sim c$).
This agrees with the linear-regime prediction derived by solving the dispersion relation in Equation~\eqref{eq:dis_cold}.
A similar pattern can be seen in the uniform regions of the ``Bump" simulation.

On the other hand, the linear regime evolution of charge density in the ``Cosine" simulation shows clear wave reflections at the higher density regions, indicated by the checkerboard patterns.
This occurs at $x \sim 90 c/\omega_p$, i.e., when the number density is about 3 times larger than that in the lowest density region (i.e., at $x = 64 c / \omega_p$).
A similar pattern (wave reflections during the linear regime evolution) is also seen in the ``Exponential" Simulation.

\begin{figure*}
\centering
\begin{tabular}{ll}
\includegraphics[width=0.49\textwidth]{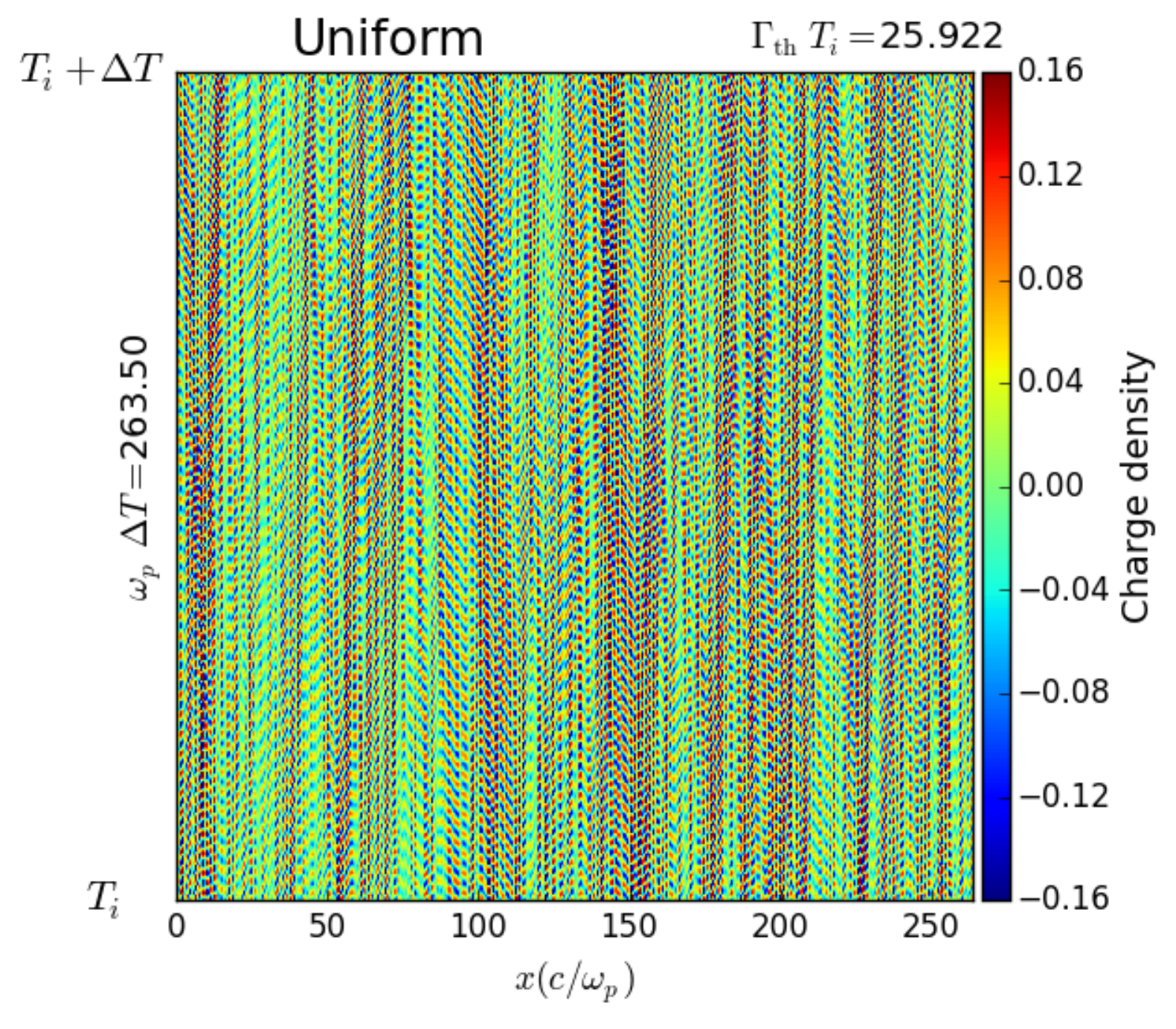}
&
\includegraphics[width=0.49\textwidth]{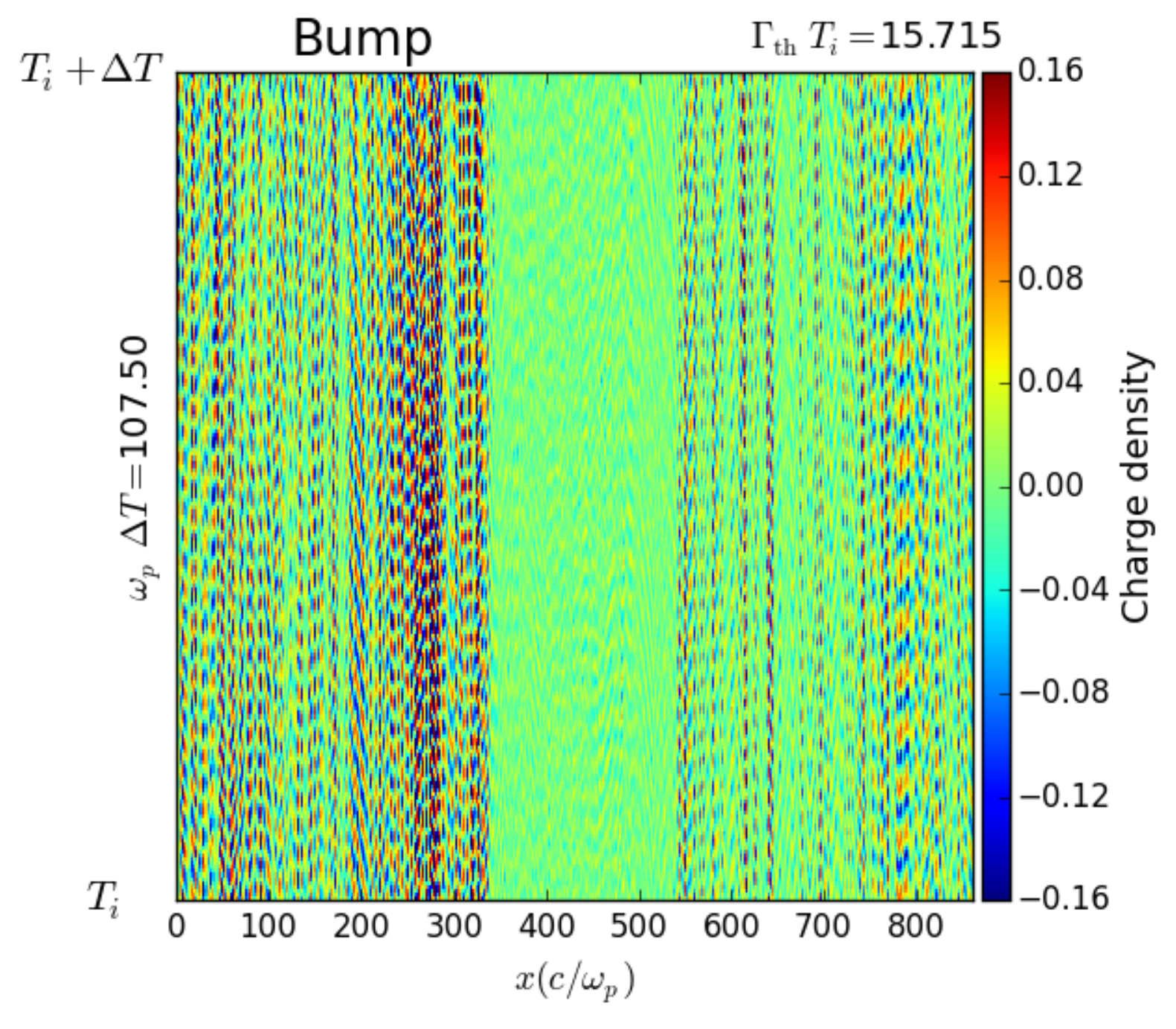}
\\
\includegraphics[width=0.49\textwidth]{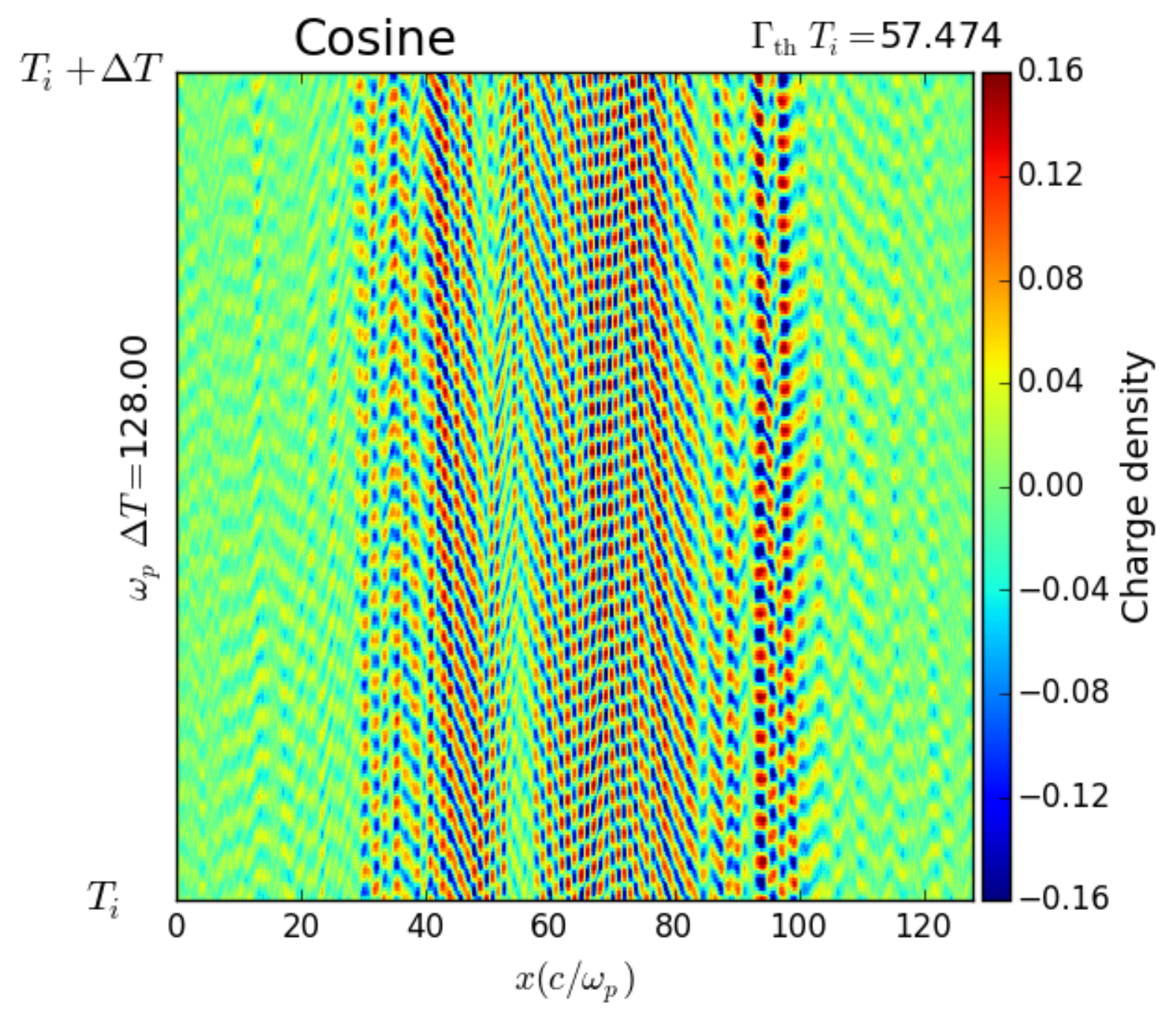}
&
\includegraphics[width=0.49\textwidth]{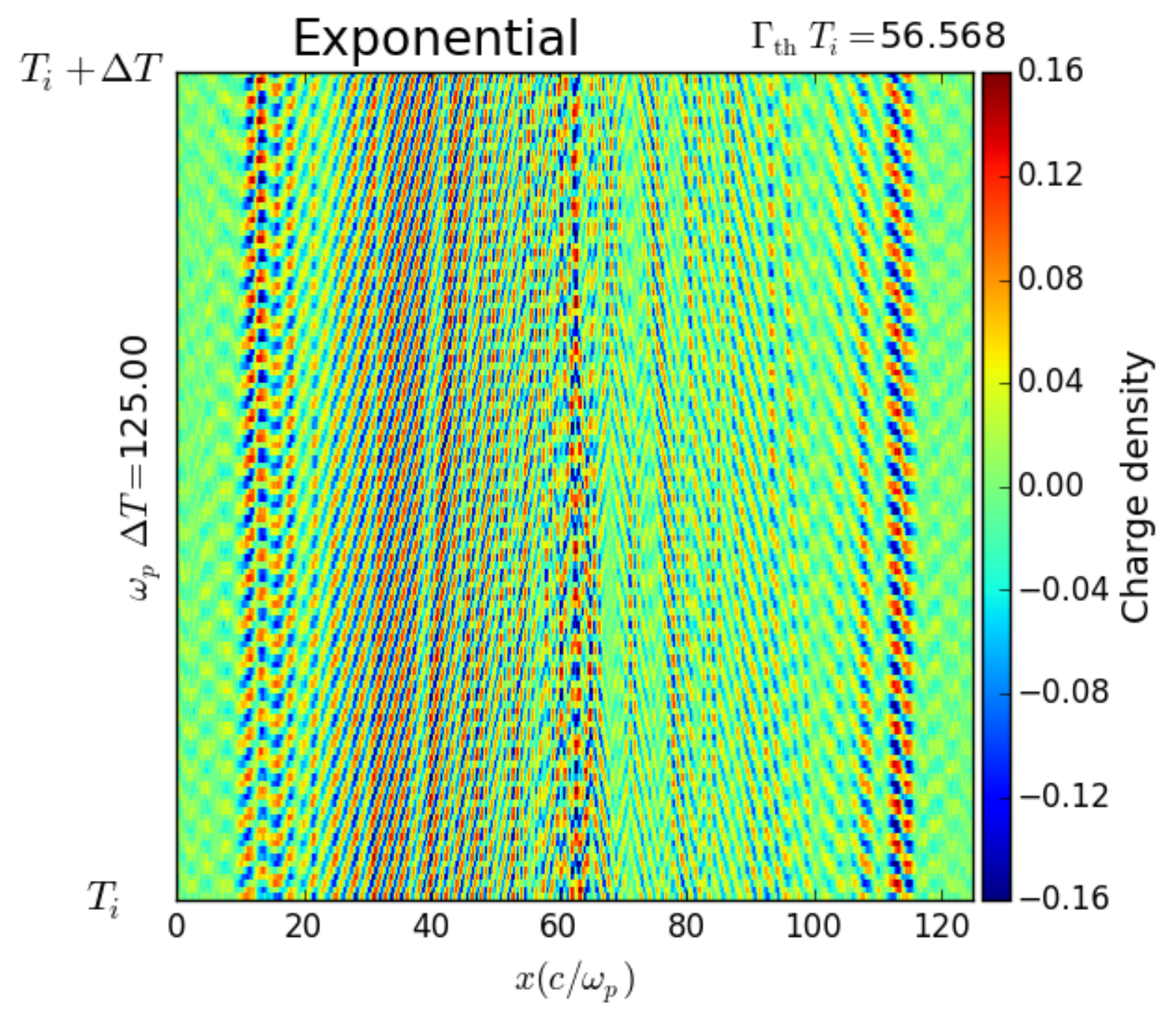}
\\
\end{tabular}
\caption{
Saturated state of the charge density in different simulations in the nonlinear regime.
The charge density is color coded as a function of position in the simulation box and and over a time period $T_i < t < T_i+\Delta T$ as indicated in the labels in each panel.
\label{fig:denxt2}
}
\end{figure*}

\subsection{Saturated non-linear regime}
\label{sec:Slevel}

\begin{figure*}
\centering
\begin{tabular}{ll}
\includegraphics[width=0.49\textwidth]{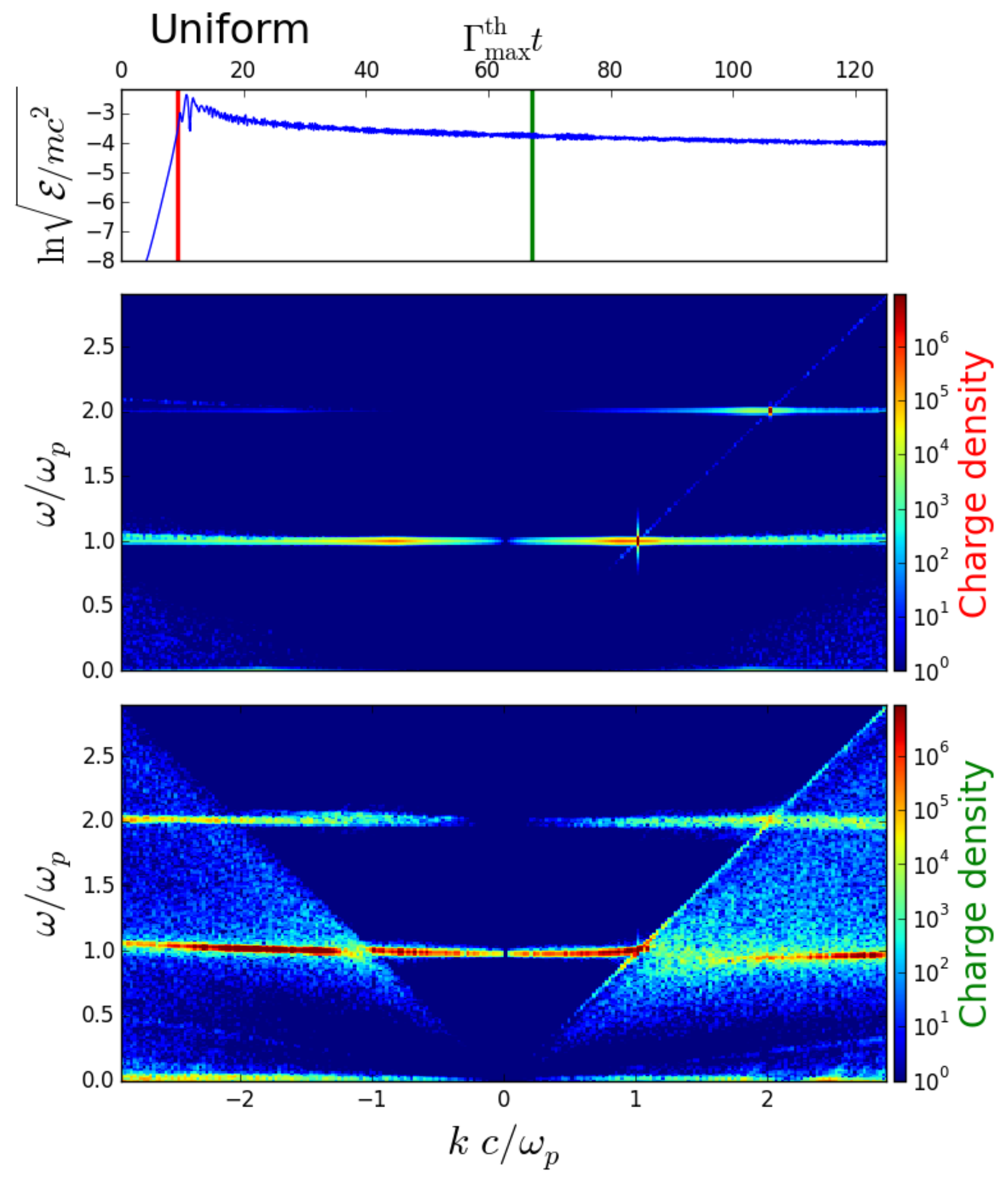}
&
\includegraphics[width=0.49\textwidth]{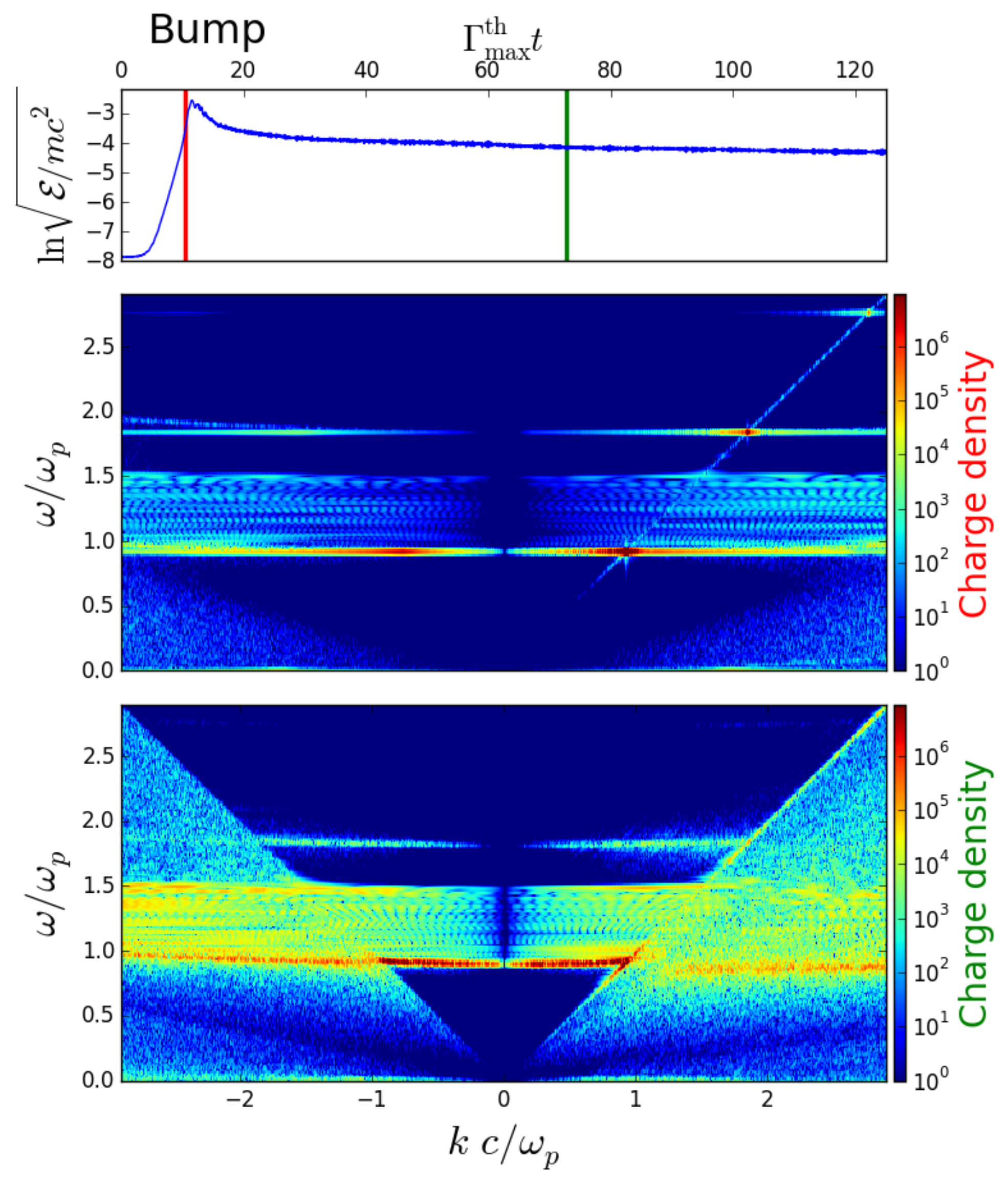}
\\
\vspace{0.2cm} &\\
\includegraphics[width=0.49\textwidth]{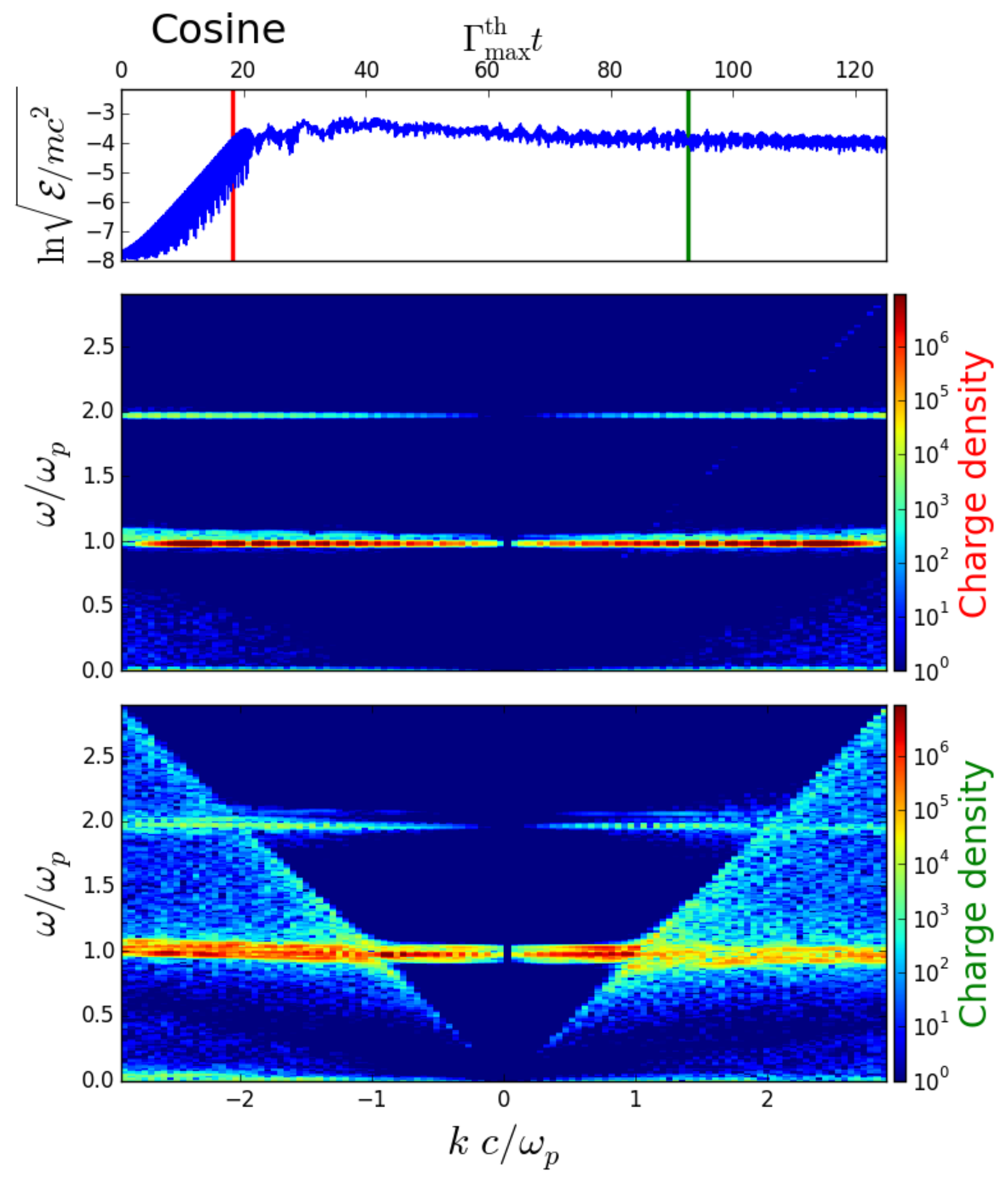}
&
\includegraphics[width=0.49\textwidth]{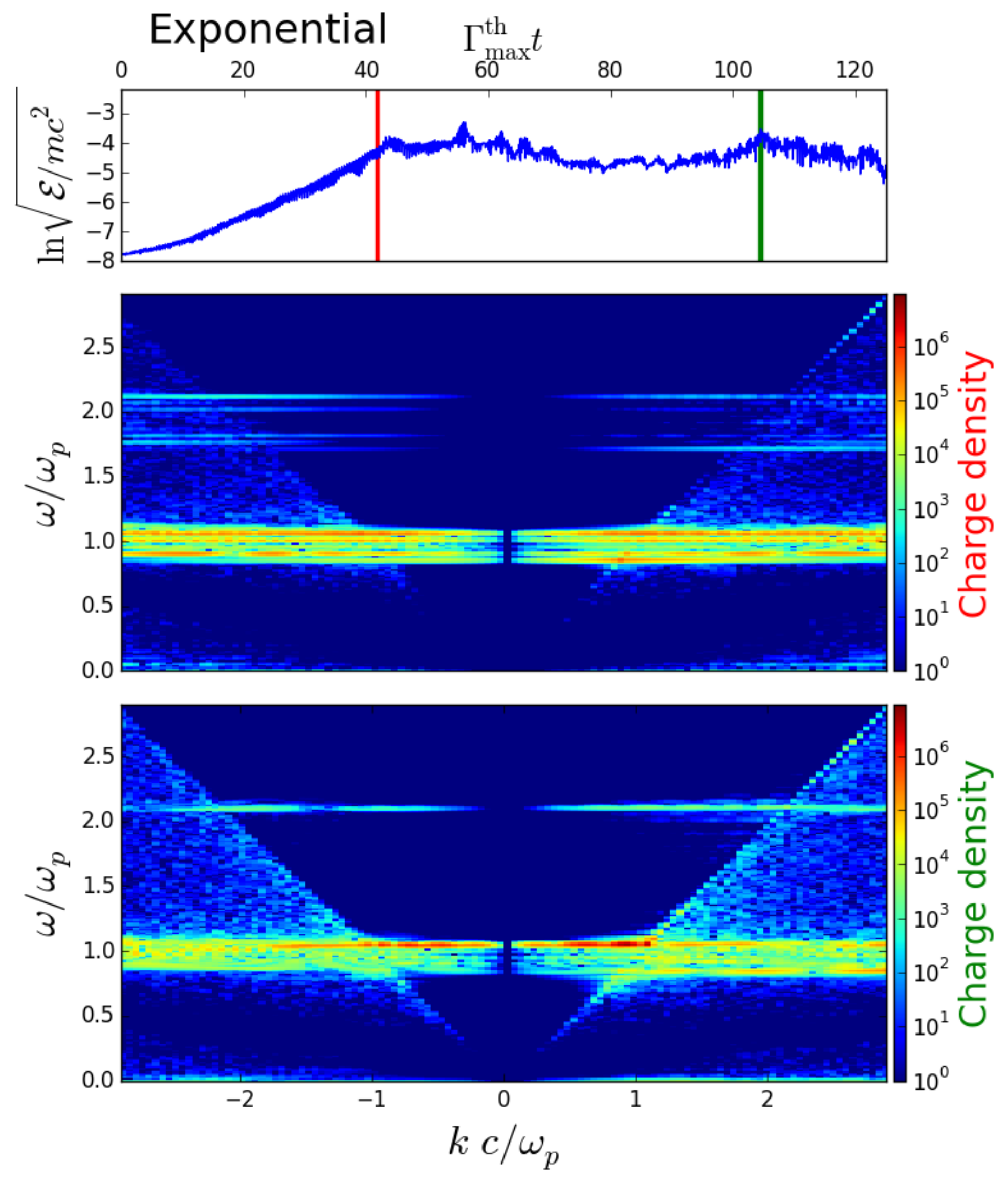}
\\
\end{tabular}
\caption{
The distribution of power (i.e., the square of the absolute value of the charge density 2D Fourier transform) of the excited beam-plasma wavemodes in all simulations.
For each simulation we show the evolution of the normalized potential energy from the linear to the saturated non-linear regime (top panel). 
The middle panel shows the distribution of power at the end of the linear evolution, as indicated by the  red vertical line in top panel.
The bottom panel shows the distribution of power in the saturated non-linear regime, as indicated by the green vertical line in the top panel.
\label{fig:denkw}
}
\end{figure*}

Analytically, it is hard to identify the reason(s) for the similar level of beam energy loss that is approximately achieved in simulations with both homogeneous and inhomogeneous background plasmas.
This is the case, despite the modestly lower growth rates (during the linear regime) of inhomogeneous plasma simulations.
Therefore, in this section we present different representations of the nonlinearly saturated state, i.e., after the end of linear growth.
The saturated state looks very similar in all simulations, which is consistent with the similar energy loss.
However, the evolution to achieve this state is different in different simulations.

In Figure~\ref{fig:denxt2}, we show the nonlinear saturated state for the charge density in all simulations.
In the nonlinear regime, the charge density evolution is complicated and clearly differs from the linear regime due to nonlinear interactions between all linearly-unstable,
growing wavemodes.
Figure~\ref{fig:denxt2} shows that the saturated state in all simulations is similar suggesting that the saturation in uniform and non-uniform simulations occur via similar physical mechanism(s) despite the difference in their evolution in the linear regime.
This is also consistent with the similar degrees of energy loss from the pair-beams in all simulations at this state.
Characterizing the physical mechanism(s) and understanding the reason(s) why they do not depend on the background plasma inhomogeneity is left for future work.

In Figure~\ref{fig:denkw}, we present the distribution of power (i.e., the square of the absolute value of the charge density 2D Fourier transform
in $(\omega,k)$-space) of different simulations at two times.
First, close to the end of the linear evolution of the grid potential energy (red), and, second, in the non-linear, saturated regime (green).
In the linear regime, the evolution of the homogeneous and inhomogeneous simulations are very different.
The presence of inhomogeneities increase the support of the unstable region in the  $(\omega,k)$-space during the linear regime.

In the linear regime, the power in both forward- and backward-propagating wavemodes in the inhomogeneous simulations is a clear sign of the reflection of forward-propagating wavemodes that are initially excited locally due to the propagation of pair beams.
In the non-linear regime, all simulations evolve to a similar physical state, where the power in linearly excited wavemodes cascades to lower $k$ (longer wavelengths).
However, in the inhomogeneous simulations the support for growing wavemodes, in $(\omega,k)$-space, is larger than that in the uniform simulations during the linear evolution and also when the nonlinear saturated state is achieved.

\section{Conclusions}
\label{sec:conclusions}

Following previous work, we derive the condition where the linear growth rates that assume a uniform background might be not applicable in presence of background inhomogeneity (Equation~\eqref{eq:inhomo-cond}).
While previous work by \citet{Miniati-Elyiv-2013} assume that violation of this condition  results in complete suppression of these beam-plasma instabilities, we demonstrated using high resolution numerical simulations that, in fact, the instability growth rates are reduced by only a factor of a few.
Moreover, the nonlinear saturation level of the instability measured in terms of the initial beam kinetic energy are broadly similar to that in the uniform background plasma case.

In the present work, the insensitivity of the level of energy loss by the pair-beam  to the background plasma inhomogeneity is explicitly demonstrated only for longitudinal unstable wavemodes.
We leave demonstrating this for oblique and perpendicular unstable wavemodes to future studies.
However, since the effect of the inhomogeneity is expected to be similar for other unstable wavemodes, our finding is likely to hold for these cases as well.

The parameters of the beam-plasma system in the IGM are extreme ($\alpha=10^{-15}$, and $\gamma_b \sim 10^6$)~\citep{blazarI}, which result in extreme separation of scales, e.g., the growth time scales are about 9 orders of magnitude longer compared to the plasma time scale ($\omega_p^{-1}$).
This makes simulating such beam-plasma systems with realistic parameters intractable.
Since the parameter in our simulations are, however, in the correct asymptotic regime, i.e., the pair-beams are subdominant in both number and energy densities compared to the background plasma ($\alpha=0.002$, and $\alpha \gamma_b = 0.2$), 
we expect our conclusions here to be directly applicable for the beam-plasma system of the IGM.

This suggests that blazar-driven beams will remain subject to virulent linear instabilities even in the presence of realistic levels of the inhomogeneity in the IGM.
The lack of the suppression of the plasma instabilities due to background plasma inhomogeneities is consistent with the lack of $\gamma$-ray halos expected around TeV blazars if plasma instabilities were suppressed~\citep{BowTiesI,BowTiesII,BowTiesIII}.

\section*{SUPPLEMENTARY MATERIAL}

Links for ($x-t$) density evolution movies:

\href{https://www.youtube.com/watch?v=tIkyl8284tE}{Cosine}
, 
\href{https://www.youtube.com/watch?v=eXWq28BrFss}{Uniform}
, 
\href{https://www.youtube.com/watch?v=zmyelWJ509I}{Bump}
, 
\href{https://www.youtube.com/watch?v=LZGAMn2sjOo}{Exponential}.

\section*{Acknowledgements}

M.S. and A.E.B.~receive financial support from the Perimeter
Institute for Theoretical Physics and the Natural Sciences and
Engineering Research Council of Canada through a Discovery Grant.
Research at Perimeter Institute is supported by the Government of
Canada through Industry Canada and by the Province of Ontario through
the Ministry of Research and Innovation.  P.C.~gratefully acknowledges
support from NSF grant AST-1255469.  C.P.~acknowledges support by the
European Research Council under ERC-CoG grant
CRAGSMAN-646955. A.L.~receives financial support from an Alfred
P. Sloan Research Fellowship, NASA ATP Grant NNX14AH35G, and NSF
Collaborative Research Grant 411920 and CAREER grant
1455342. E.P.~acknowledges support by the Kavli Foundation.

\bibliography{refs_01,refs_02,refs_03}
\bibliographystyle{apj}

\end{document}